\documentclass[amsmath,amssymb,aps,prl,twocolumn, superscriptaddress,longbibliography]{revtex4-2}
\usepackage{graphicx}
\usepackage{dcolumn}
\usepackage{xcolor}
\usepackage{helvet}
\usepackage{mathrsfs}
\usepackage[colorlinks=true,bookmarks=false,citecolor=blue,urlcolor=blue]{hyperref}
\usepackage{physics}
\usepackage{dsfont}

\renewcommand{\cite}[1]{[\onlinecite{#1}]}
\newcommand{\cra}[1]{\hat{a}^{\dag}_{#1}}  
\newcommand{\ana}[1]{\hat{a}_{#1}}         

\begin{document}

\title{Time-optimal transfer of the quantum state in long qubit arrays}

\author{Andrei~A.~Stepanenko}
\email{as@lims.ac.uk}
\affiliation{London Institute for Mathematical Sciences, Royal Institution, London, UK}
\affiliation{School of Physics and Engineering, ITMO University, Saint Petersburg 197101, Russia}

\author{Kseniia S. Chernova}
\affiliation{School of Physics and Engineering, ITMO University, Saint Petersburg 197101, Russia}

\author{Maxim~A.~Gorlach}
\email{m.gorlach@metalab.ifmo.ru}
\affiliation{School of Physics and Engineering, ITMO University, Saint Petersburg 197101, Russia}

\begin{abstract}
Recent technological advances have allowed the fabrication of large arrays of coupled qubits serving as prototypes of quantum processors. However, the optimal control of such systems is notoriously hard, which limits the potential of large-scale quantum systems. Here, we investigate a model problem of quantum state transfer in a large nearest-neighbor-coupled qubit array and derive an optimal control that simultaneously enables maximal fidelity and minimal time of the transfer.   
\end{abstract}

\maketitle


{\it Introduction.~--} Rapid progress in quantum technologies enabled large-scale quantum systems capable of performing quantum algorithms and quantum simulations. Existing platforms include trapped ions~\cite{Bruzewicz2019,Moses2023,Chen2024}, cold atoms~\cite{Lukin2023}, photonic systems~\cite{Madsen2022,Taballione2023} as well as arrays of superconducting qubits~\cite{Arute2019,JianWeiPan2021,Bravyi2022,Anferov2024,Morvan2024,GoogleQuantumAI}. Recent years have witnessed rapid growth in the capabilities of such noisy intermediate-scale quantum systems~\cite{Preskill2018,Bharti2022} and an active discussion of the quantum supremacy concept~\cite{Arute2019,JianWeiPan2021,Pan2022,Bulmer2022,GoogleQuantumAI}.


To fully harness the scales of modern quantum systems, it is important to have their complete and flexible control. Therefore, the strategies of quantum optimal control~\cite{Werschnik2007,Boscain2021} are under active investigation. Popular approaches include counter-adiabatic driving~\cite{Demirplak2003,Berry2009} and shortcuts to adiabaticity~\cite{Campo2013,GueryOdelin2019} as well as quantum brachistochrone method~\cite{Carlini2006,Carlini2007,Wang2015,Malikis2024} based on the geometric approach~\cite{Wang2015,Meinersen2024}.


The latter technique is based on a variational formulation aiming to maximize the speed of the transition (i.e. minimize its time duration) given the constraints on the system Hamiltonian. This approach has a geometric interpretation in terms of geodesics~\cite{Wang2015} and produces a tractable system of differential equations~\cite{Carlini2007} that can be solved analytically for relatively simple quantum systems with few degrees of freedom. However, the direct application of this or any other method to large-scale quantum systems is challenging because of the overwhelming number of degrees of freedom and extremely large parameter space.

\begin{figure}[b]
    \centerline{
    \includegraphics[width = 0.49\textwidth]{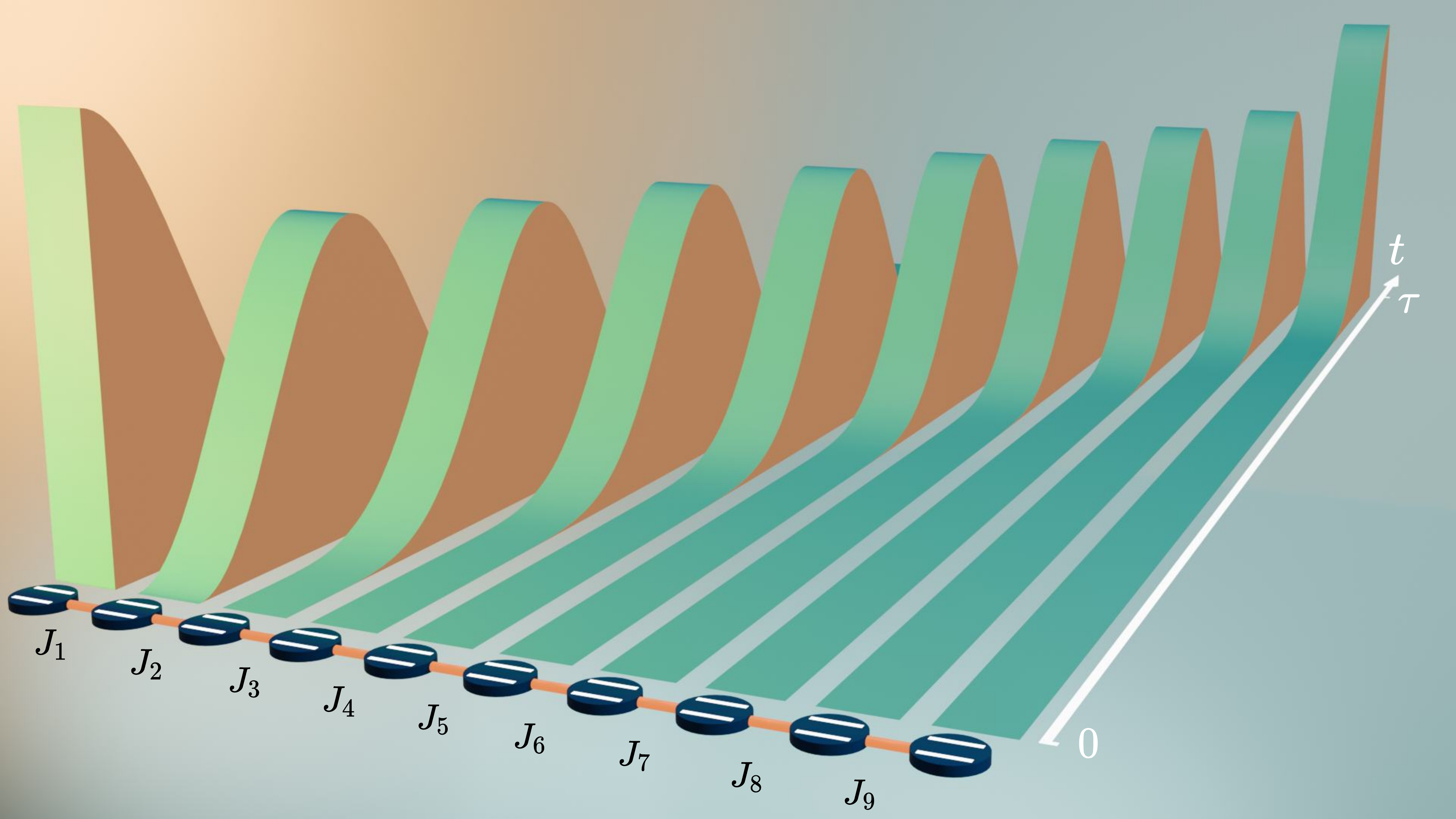}}
    \caption{An artistic view of single-particle transfer under the time-optimal control of couplings $J_m(t)$ in an array of $N=10$ coupled qubits. The qubits are depicted by the dark blue cylinders.   
    }
    \label{fig:1}
\end{figure}

In this Letter, we make a conceptual step to solve this problem. As a physically motivated example, we consider an array of $N$ nearest-neighbor-coupled qubits assuming that the couplings $J_n(t)$ between them can be tailored on demand and controlled in real time so that the overall sum $\sum_n\,J_n^2(t)$ is bounded by the constant value $J_0^2$ [Fig.~\ref{fig:1}]. Though challenging, such real-time control of couplings is technically feasible. For instance, in superconducting architecture this could be achieved by inserting auxiliary off-resonant qubits; the change of their eigenfrequencies will renormalize the effective couplings $J_m$~\cite{Keil2016}.
%
For clarity, we also neglect the effects of dissipation and decoherence, focusing on the control of a Hermitian system. As the simplest protocol, we consider the transfer of a single excitation initiated in the first (leftmost) qubit to the $N^{\text{th}}$ (rightmost) position. With a suitable modification of the quantum brachistochrone approach, we derive the optimal control of this system, enabling simultaneously maximal fidelity of the transfer along with the minimal transfer time. Interestingly, the transfer time scales {\it linearly} with the length $N$ of the array, which correlates with the intuitive picture of a Gaussian-type wavepacket travelling in the array with the highest possible speed.


We model the array of qubits with a Hamiltonian 
\begin{equation}\label{eq1:Hamiltonian}
\hat{H} = \sum_m J_{m}(t)\,\left(\cra{m}\ana{m+1} +\cra{m+1}\ana{m}\right)\;,
\end{equation}
where $\ana{m}$ is an annihilation operator in the $m^{\text{th}}$ qubit. We also assume that the eigenfrequencies of all qubits are identical and hence the contribution $\sum_m\,\omega_m\,\cra{m}\ana{m}$ results only in a constant energy shift which does not affect the transfer process and is omitted in Eq.~\eqref{eq1:Hamiltonian} for clarity.

The Hamiltonian Eq.~\eqref{eq1:Hamiltonian} commutes with a total number of excitations $\hat{n}=\sum_m\,\cra{m}\,\ana{m}$:  $\left[\hat{H},\hat{n}\right]=0$. Hence, the number of excitations is conserved and the $N$-dimensional single-particle sector spanned by the basis vectors $\ket{m}=\cra{m}\,\ket{0}$ is decoupled from the entire $2^N$-dimensional Hilbert space of the system. This allows us reduce the complexity of the problem and present the Hamiltonian as $N\times N$ Hermitian matrix.



{\it Quantum brachistochrone method and governing equations.~--} To find an optimal strategy to switch the couplings $J_m(t)$, we adapt quantum brachistochrone method~\cite{Carlini2006, Carlini2007, Wang2015, Malikis2024}. We introduce the evolution operator $\hat{U}(t)$ which connects the states of the system at two distinct moments of time as $\ket{\psi(t)}=\hat{U}(t)\,\ket{\psi(0)}$ and satisfies Shr{\"o}dinger equation
\begin{equation}
i\,\partial_t\hat{U} = \hat{H}(t)\hat{U}\:.    
\end{equation}
Since $\text{Tr}\,\hat{H}=0$ at all times, the Hamiltonian of the array belongs to $\left(N^2-1\right)$-dimensional space of zero-trace Hermitian matrices. $(N-1)$ of those matrices $\hat{A}_n = (\cra{n}\ana{n+1} +\cra{n+1}\ana{n})/\sqrt{2}$ describe the coupling of the neighboring sites spanning the subspace $\mathcal{A}$, while the remaining $(N^2-N)$ matrices $\hat{B}_l\in \mathcal{B}$ are responsible for the long-range interactions absent in our system. The introduced matrices are normalized by the conditions $\text{Tr}\,(\hat{A}_m\,\hat{A}_n)=\delta_{mn}$, $\text{Tr}\,(\hat{B}_k\,\hat{B}_l)=\delta_{kl}$, $\text{Tr}\,(\hat{A}_m\,\hat{B}_k)=0$.

We aim to minimize the time of the transfer $\tau$ given a fixed bound $J_0^2$ on the sum of squares of the couplings. Although this can be done directly~\cite{Carlini2006,Carlini2007,Yang2022}, there is an equivalent, but simpler formulation aiming to minimize $J_0=\sqrt{\sum_n J_n^2}$ for the fixed time of the transfer $\tau$. In turn, the sum $\sum_n J_n^2$ can be recast as $\text{Tr}\,\hat{H}^2\equiv ||\hat{H}||^2$, resulting in the cost functional $S_0=\int\limits_0^\tau\,||\hat{H}(t)||\,dt$.

If the Hamiltonian is unrestricted, the best possible strategy is to couple the initial $\ket{\psi_0}$ and target $\ket{\psi_1}$ state directly by the maximal possible coupling~\cite{Carlini2006}. In our case, however, the Hamiltonian includes only nearest-neighbor couplings which prevents the direct transfer from the first to $N^{\text{th}}$ qubit. Therefore, we introduce an additional contribution $S_1=\int_0^\tau\,\mathrm{Tr} (\hat{D}\hat{H})\,dt$, where $\hat{D} = \sum_l \lambda_l\hat{B}_l$ contains the matrices from $\mathcal{B}$ subspace and $\lambda_l$ are time-dependent Lagrange multipliers ensuring that the Hamiltonian at any moment of time does not contain any of $\hat{B}_l$ matrices: $\text{Tr}\,(\hat{H}\,\hat{B}_l)=0$.

Finally, we add two boundary terms to ensure that the state $\ket{\psi_0}$ in the initial moment $t=0$ is transferred to the target state $\ket{\psi_1}\,e^{i\phi}$ at $t=\tau$, where the global phase $\phi$ is irrelevant. Hence, the overall cost functional takes the form
\begin{eqnarray}
\label{eq2:action}
    S &=&  \int_{0}^{\tau}\,||\hat{H}(t)||\,dt
    + \int_{0}^{\tau} \mathrm{Tr}\left(\hat{D}\hat{H}\right)\,dt\nonumber\\
    &&+\mathrm{Tr}\left(\hat{R}_0(\hat{U} \hat{P}_0\hat{U}^\dagger-\hat{P}_0)\right)\delta(t)\nonumber\\
    &&+\mathrm{Tr}\left(\hat{R}_1(\hat{U} \hat{P}_0\hat{U}^\dagger-\hat{P}_1)\right)\delta(t-\tau)\;,
\end{eqnarray}
%
where $\hat{R}_0$ and $\hat{R}_1$ are the matrices of Lagrange multipliers, and $\hat{P}_{0,1}$ projectors project on initial and target states of the system as  $\hat{P}_0 = \ket{\psi_0}\bra{\psi_0}$, $\hat{P}_1 = \ket{\psi_1}\bra{\psi_1}$. 

Since the Hamiltonian is expressed in terms of $\hat{U}$ as $\hat{H}=i\,(\partial_t\hat{U})\,\hat{U}^\dagger$, the above functional depends on the evolution operator $\hat{U}(t)$, its time derivative and Lagrange multipliers $\lambda_l$, $\hat{R}_0$, $\hat{R}_1$. Varying with respect to $\hat{U}$ and requiring $\delta S=0$, we derive {\it quantum brachistochrone equation}~\cite{Carlini2007,Wang2015,Malikis2024}
\begin{eqnarray}
    \label{eq3:QBmat}
    \partial_t(\hat{H} + \hat{D})+i\left[\hat{H},\hat{D}\right] = 0\:,
\end{eqnarray}
which defines the change of the Hamiltonian in time. An immediate consequence of Eq.~\eqref{eq3:QBmat} is
\begin{equation}\label{eq:TrD}
    \text{Tr}\,\hat{D}=\text{const}\:,
\end{equation}
where the constant on the right-hand side is determined by the initial conditions.

However, finding the optimal protocol from Eq.~\eqref{eq3:QBmat} is generally a challenging task, since the initial conditions for the couplings $J_m(0)$ and Lagrange multipliers $\lambda_l(0)$ are unknown. For small-scale quantum systems, this issue can be overcome by defining the evolution operator in the initial and final moments of time and solving the resulting boundary value problem either analytically or by the shooting method~\cite{Carlini2007}. Further improvement is obtained by connecting the solutions of Eq.~\eqref{eq3:QBmat} to the geodesics in the space with a special metric~\cite{Wang2015}. Here, we pursue a different route and derive the boundary conditions varying the two terms of $S$ with the delta-function:
\begin{eqnarray}
\label{eq4:bound0}
    \hat{H}(0)
    + \hat{D}(0)
     &=& -i\left[\hat{P}_0,\hat{R}_0\right]\;,\\
\label{eq5:bound1}
\hat{H}(\tau)
    + \hat{D}(\tau)
     &=& i\left[\hat{P}_1,\hat{R}_1\right] \;.
\end{eqnarray}

Some of the scalar equations in the system \eqref{eq4:bound0}-\eqref{eq5:bound1} are independent of $\hat{R}_0$ and $\hat{R}_1$ components, and those provide the boundary conditions of interest. For our choice of initial and final states,
\begin{eqnarray}
  \label{eq:BC1fin}
    H_{ij}(0)
    + D_{ij}(0)
     &=& 0\;,\\
\label{eq:BC2fin}
H_{kl}(\tau)
    + D_{kl}(\tau)
     &=& 0 \;,\\
\label{eq:BC3fin}
\text{Tr}\,\hat{D}(0)=\text{Tr}\,\hat{D}(\tau)&=&0\:,
\end{eqnarray}
where $i,j=2,3,\dots,N$ and $k,l=1,2,\dots, N-1$. This formulation of quantum brachistochrone Eqs.~\eqref{eq3:QBmat}-\eqref{eq5:bound1} significantly simplifies the calculation, as we have to seek not the entire matrix of the evolution operator with $N^2$ components, but rather the wave function $\ket{\psi}$ with only $N$ entries solving the Schr{\"o}dinger equation
\begin{equation}\label{eq:WaveFuncEq}
    i\frac{\partial\ket{\psi}}{\partial t}=\hat{H}\,\ket{\psi}\:.
\end{equation}
with $2N-1$ initial and boundary conditions for the wave function:
\begin{equation}\label{eq:BCWaveF}
    \psi_n(0)=\delta_{n1},\mspace{8mu} \psi_k(\tau)=0\:,
\end{equation}
where $n=1,\dots,N$ and $k=1,\dots, N-1$.

{\it Control algorithm.~--} To proceed with the solution, we choose a specific basis in the $(N^2-1)$-dimensional space of traceless Hermitian matrices. To that end, we introduce a matrix function
\begin{equation}
    \hat{X}_{mn}(z)=\frac{1}{\sqrt{2}}\,\left(z\,E_{nm}+z^*\,E_{mn}\right)\:,
\end{equation}
where $z$ is an arbitrary complex number and $E_{nm}$ is a matrix with the elements $(E_{nm})_{pq}=\delta_{np}\,\delta_{mq}$. In these notations, $\hat{A}_m=\hat{X}_{m,m+1}(1)$. In turn, the matrices from the $\mathcal{B}$ subspace are constructed as $\hat{B}_{m,m+q}^e=\hat{X}_{m,m+q}(i^{q-1})$ for $1<q\leq N-m$, $\hat{B}_{m,m+q}^o=\hat{X}_{m,m+q}(i^{q})$ for $1\leq q\leq N-m$, and $\hat{B}_{m,m}=(\sum_p^m E_{pp}-m\,E_{m+1,m+1})\sqrt{2/(m^2+m)}$ where $1\leq m <N$. This choice of the matrices provides slight modification of generalized Gell-Mann matrices~\cite{Bertlmann2008}.

In such basis, quantum brachistochrone equation Eq.~\eqref{eq3:QBmat} results in a set of scalar equations (see Supplementary Materials) 
\begin{eqnarray}
\label{eq6:Jctrl}
    \sqrt{2}\,\partial_t J_{m}&=& J_{m+1}\lambda_{m,m+2}-J_{m-1}\lambda_{m-1,m+1}\;,\\
\label{eq7:Lctrl}
    \partial_t\lambda_{k,k+n} &=& 
    J_{k+n} \lambda_{k,k+n+1} 
    -J_{k-1} \lambda_{k-1,k+n}\nonumber\\&&
    - J_{k+n-1} \lambda_{k,k+n-1}
    + J_{k} \lambda_{k+1,k+n}\;,
\end{eqnarray}
where $1<n\leq N-k$ and $1\leq k < N$. Notably, the system only contains Lagrange multipliers corresponding to the matrices $\hat{B}_{m,m+q}^e$, while the terms associated with $\hat{B}_{m,m+q}^o$ and $\hat{B}_{m,m}$ drop out due to the structure of the problem.

Applying the boundary conditions Eqs.~\eqref{eq:BC1fin}-\eqref{eq:BC3fin}, we recover that most of the couplings at the initial moment are zero $J_m(0) = 0$ for $m\neq 1$, while at the final moment $J_m(\tau) = 0$ for $m\neq N-1$. This result is very intuitive: to transfer the excitation from the first qubit elsewhere one has to maximize the coupling $J_1$ keeping the rest of the couplings zero.

What is less intuitive, the major part of the coefficients $\lambda_{k,k+n}$ is also zero at $t=0$ and $t=\tau$ with the only nonzero coefficients $\lambda_{1,2+n}(0)$ and $\lambda_{n,N}(\tau)$ for $1\leq n<N-2$.



Thus, our problem has $N(N+1)/2$ unknowns including $N$ complex components of the wave function $\ket{\psi}$, $(N-1)$ real couplings $J_m$ and $(N-1)(N-2)/2$ real-valued Lagrange multipliers appearing in the problem. They satisfy the same number of  independent differential equations~\eqref{eq3:QBmat}, \eqref{eq:WaveFuncEq}.

This is supplemented by $(N^2-N+1)$ initial and boundary conditions including $(2N-1)$ conditions for the wave function Eqs.~\eqref{eq:BCWaveF} and $(N-1)(N-2)$ conditions for quantum brachistochrone equation (see Supplementary Materials). Hence, starting from $N=3$, the number of conditions exceeds the number of equations and the system becomes overdetermined. Physically, this reflects the fact that the optimal solution may not exist. However, as we demonstrate below, the optimal solution exists and is uniquely constructed.

Specifically, we determine $J_1(0)$ and $\lambda_{1,2+n}(0)$ by the shooting method. Importantly, our formulation requires only $(N-1)$ initial conditions for the shooting. As a result, the numerical search for short arrays ($N<5$) converges for practically random initial guess.

The computation is less trivial for longer arrays. In this case, however, we proceed iteratively using the solution for the array with $N-1$ qubits to construct  the initial guess for $N$-qubit problem. The standard shooting method works relatively well on a personal computer up to $N<17$. For longer arrays, we seek $J_1(0)$ and $\lambda_{1,1+q}(0)$ by the gradient optimization method yielding the solution for $N$ as large as 100.

\begin{figure}[b]
    \centerline{
    \includegraphics[width = 0.49\textwidth]{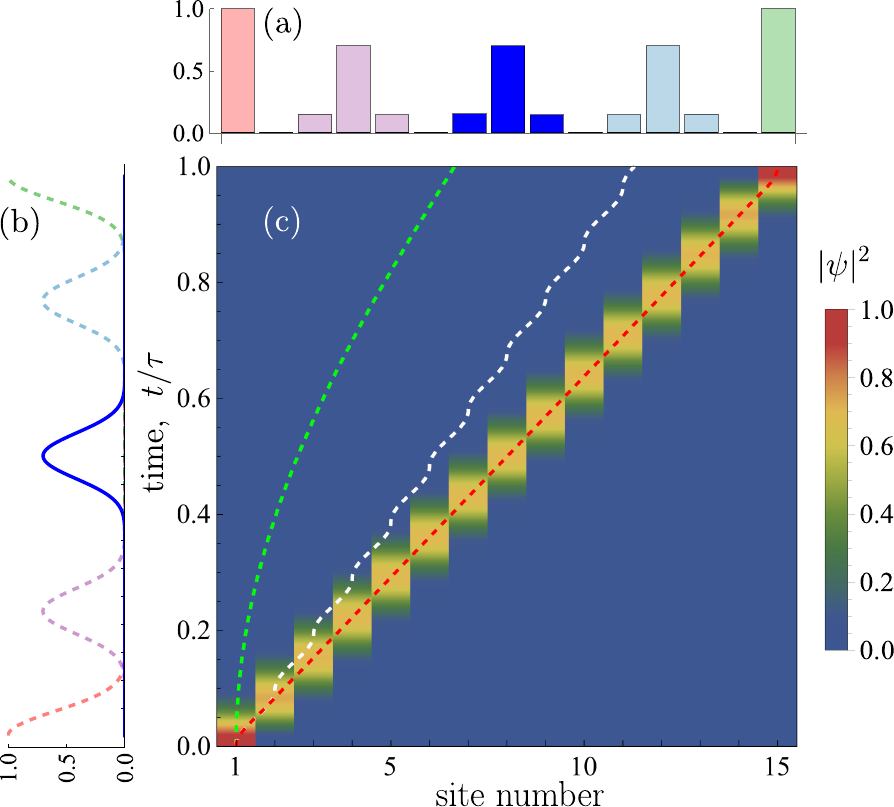}}
    \caption{Numerical results for time-optimal transfer in a 15-qubit array. (a) Histogram showing the instantaneous probability distribution $|\psi_n(t)|^2$ at several representative moments of time. The wavepacket remains tightly bound. (b) The dependence of probabilities $|\psi_n(t)|^2$ on time in several selected sites of the lattice. (c) Evolution of the probability distribution in the array during the whole process of the transfer $0\leq t \leq \tau$. Lines show the expectation value of the wavepacket position for time-optimal control (red), stepwise switching of the couplings (white) and ``perfect transfer'' scenario (green line).}
    \label{fig:2}
\end{figure}

{\it Key results.~--} To illustrate our approach, we compute the optimal control and the associated evolution of the wave function for the array consisting of $15$ qubits. Figure~\ref{fig:2}(a) depicts the probability distribution $|\psi_n|^2$ for the quantum state at several fixed moments of time. We observe that the excitation propagates in the array as a tightly bound wave packet retaining its shape with a modification happening only close to the boundaries.

On the other hand, tracking the evolution of probabilities $|\psi_n|^2$ versus  time for various site numbers $n$ [Fig.~\ref{fig:2}(b)], we observe that the curves peak as the wavepacket passes the respective site, while the curves for the different site index $n$ are practically identical up to the shift in time.


Finally, Fig.~\ref{fig:2}(c) shows the evolution of the quantum state both in space and time. For clarity, we compare the derived time-optimal strategy with the two alternative scenarios ensuring maximal fidelity of the transfer.

The first approach is a stepwise switching of the couplings. At each time step of duration $\Delta\tau=\pi/(2\,J_0)$, only two qubits are coupled with each other. Switching the couplings one after another this way, one can transfer the excitation from the first to the $N^{\text{th}}$ qubit within the time 
\begin{equation}\label{eq:Step}
  \tau_{st} = \frac{(N-1)\pi}{2\,J_0}.  
\end{equation}
The expectation value of the particle position versus time in this case is shown in Fig.~\ref{fig:2}(c) by the dashed white line.

Another strategy is to keep all couplings in the array constant in time, but dependent on coordinate: $J_m =\gamma\sqrt{m(N-m)}/2$. Such scenario called perfect transfer~\cite{Christandl2004} also ensures maximal fidelity. However, the timing here is clearly non-optimal 
\begin{equation}\label{eq:Perfect}
   \tau_{p}=\pi/\gamma = \frac{\pi}{J_0}\,\sqrt{\frac{N(N^2-1)}{24}} 
\end{equation}
and scales roughly as $N^{3/2}$. This scenario is illustrated in Fig.~\ref{fig:2}(c) as a dashed green line. Notably, both scenarios are significantly slower compared to our time-optimal solution [Fig.~\ref{fig:2}(c)].

\begin{figure}[b]
    \centerline{
    \includegraphics[width = 0.49\textwidth]{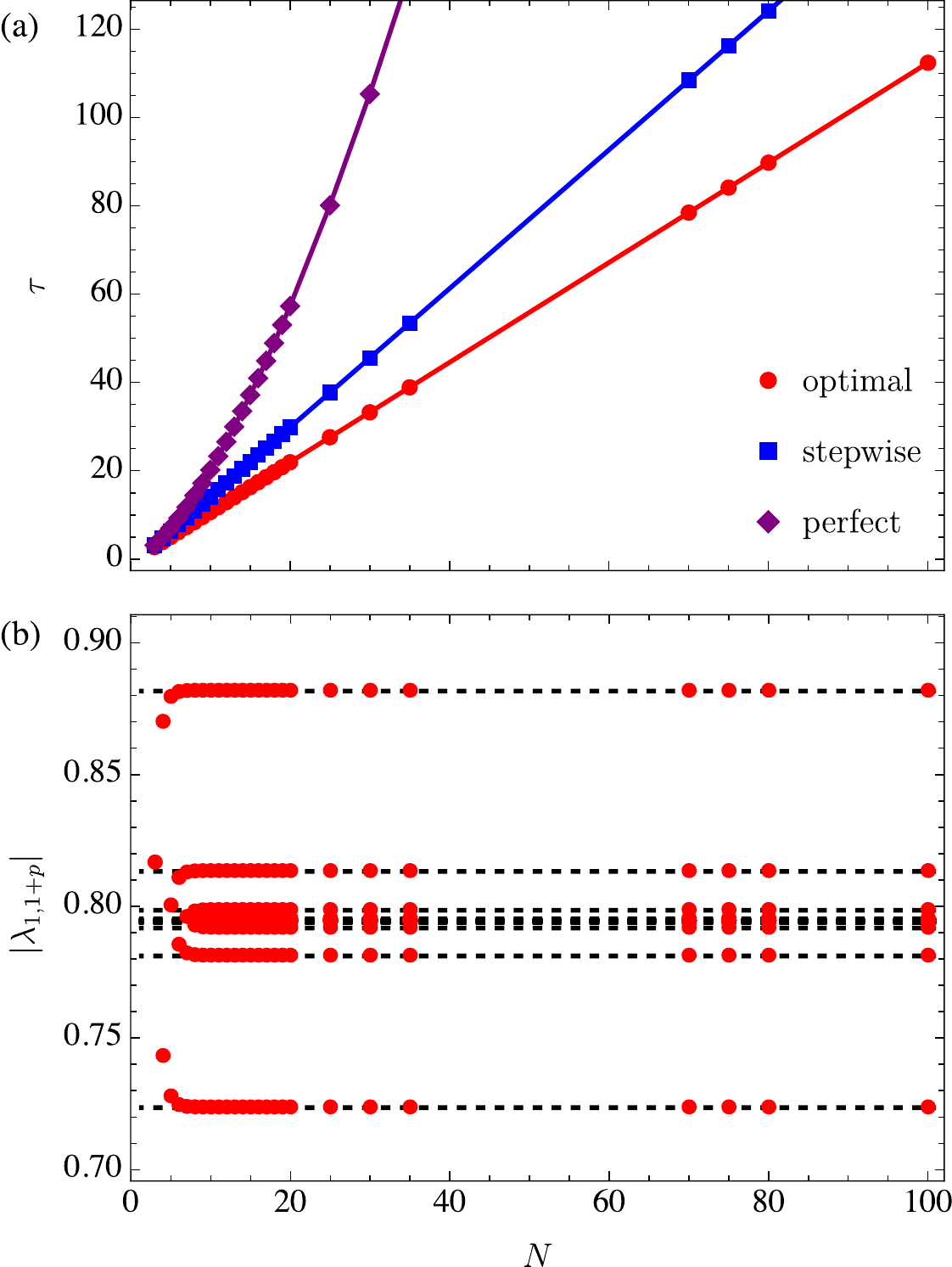}}
    \caption{Scaling of the transfer time and parameters of numerical solution with the length $N$ of the array. (a) Transfer time versus the length of the array for time-optimal control (circles), stepwise switching of the couplings (squares) and ``perfect transfer'' (rhombs). (b) Initial values of $\lambda_{1,1+p}$ parameters needed for numerical solution. Black dashed line shows projected asymptotic values in the limit $N\rightarrow\infty$.}
    \label{fig:3}
\end{figure}

Having an efficient numerical procedure to solve quantum brachistochrone equations, we now analyze the scaling of the transfer time with the length $N$ of the array. In our calculations, we examine sufficiently large qubit arrays with $N$ reaching state-of-the-art levels of 100~\cite{GoogleQuantumAI}. Our results suggest that the transfer time scales {\it linearly} with the length $N$ which agrees with the intuitive picture of a wavepacket propagating with the maximal possible speed retaining its spatial profile. The dependence of the transfer time on $N$ is well approximated by
\begin{equation}\label{eq:Scaling}
   \tau(N) = \left(1.13045\,N-0.6677\right)/J_0\:.
\end{equation}
%
This asymptotic formula is valid for sufficiently large $N$ and slightly underestimates the transfer time having an absolute error of $0.0003$ for $N=10$.


For clarity, we compare these results with the two alternative approaches summarized above. Stepwise switching of the couplings [Eq.~\eqref{eq:Step}] also provides a linear scaling, but the time of the transfer in the limit $N\rightarrow\infty$ is $39\%$  higher than for our solution. Even poorer results are obtained for time-independent couplings (``perfect transfer'' scenario), when the transfer time Eq.~\eqref{eq:Perfect} scales as $N^{3/2}$.

This comparison highlights the potential of optimal control which appears to be especially fruitful for large-scale quantum systems providing an optimization of such standard task as quantum state transfer.

However, finding the optimal control for large quantum systems is not always straightforward. In our case, a clue to the efficient numerical solution is provided by the asymptotic behavior of $\lambda_{1,1+p}$ coefficients, which remain practically constant once $N$ becomes larger than 10 [Fig.~\ref{fig:3}(b)]. Hence, having a solution for the array of $N$ qubits, we can immediately construct good initial guess for a longer array improving it by the gradient optimization.

To conclude, our work derives an example of time-optimal control for the large-scale array of nearest-neighbor-coupled qubits. Despite its conceptual simplicity, our model embodies the features of the present-day superconducting quantum processors and demonstrates the ways to significantly boost their performance by utilizing optimal control strategy. As we prove, quantum brachistochrone technique combined with the suitable numerical algorithms provides a significant speed-up of quantum state transfer as compared to more traditional approaches.

We believe that this study may stimulate further advances in time-optimal preparation and transfer of various quantum states as well as optimization of next-generation quantum algorithms.


\begin{acknowledgments}
We acknowledge D. Stepanenko for the help with illustrations. We thank O.~Gamayun, A.~Fedorov, S.~Kilin, D.~Mogilevtsev and A.~Mikhalychev for valuable discussions. 
M.A.G. acknowledges partial support by the Foundation for the Advancement of Theoretical Physics and Mathematics ``Basis''.
\end{acknowledgments}

\bibliography{article/ref}

\clearpage 
\onecolumngrid
\section*{Supplementary Materials} 
\setcounter{section}{0} 
\setcounter{equation}{0} 
\renewcommand{\thesection}{S\arabic{section}} 
\renewcommand{\theequation}{S\arabic{equation}} 
\renewcommand{\thefigure}{S\arabic{figure}} 
\section{I. Derivation of quantum brachistochrone equation}

Below, we derive quantum brachistochrone equation in the most compact form together with the boundary conditions used for the numerical solution.

We assume that the matrices $\hat{M}_m \in \mathcal{M}$ form a full orthonormal basis normalized by the condition $\mathrm{Tr}\left(\hat{M}_m\hat{M}_n\right)=\delta_{mn}$. These matrices can be viewed as generators of $SU(N)$ group. Consider the evolution of the quantum state governed by the Shr\"odinger equation $i\partial_t\hat{U} = \hat{H}(t)\hat{U}$, with a unitary evolution operator $\hat{U}(t)$ and Hamiltonian $\hat{H}(t) = \sqrt{2}\,\sum_{m} J_{m}(t)\hat{A}_m$, where $\hat{A}_m \in \mathcal{A}$ is a subset of $\mathcal{M}$ and $\hat{A}_m$ matrices are defined in the main text. 

We consider the Hamiltonian with a bounded norm $||\hat{H}(t)||\leq \mathcal{E}$. Note that the time variable can always can be rescaled $t\rightarrow g(t)$, where $g(t)$ is a monotonically increasing function. Thus, by choosing  $\partial_t g(t) = ||H(t)||/\mathcal{E}$ we find that the evolution follows along the same trajectory $i\partial_t\hat{U}(t)=\hat{H}\hat{U} \Rightarrow i\partial_g\hat{U}(t)=\hat{H}_g\hat{U}$, where $\hat{H}_g = \mathcal{E}\hat{H}(t)/||\hat{H}(t)||$ is a Hamiltonian with a constant and maximal norm $||\hat{H}_g(t)|| = \mathcal{E}$. Therefore, to simplify our analysis further, we consider only Hamiltonian with the constant norm $H(t)\equiv H_g(t)$.

To find $\hat{U}(t)$ which ensures the transfer of a quantum state in a minimal possible time $\tau$, we consider the following cost functional
\begin{eqnarray}
    S = S_1 + S_2 + S_3\,
\end{eqnarray}
where the first term minimizes the norm of the Hamiltonian $||\hat{H}|| = \sqrt{\mathrm{Tr}(\hat{H}^\dagger \hat{H})}$
\begin{eqnarray}
    S_1 = \int_{0}^{\tau}dt||\hat{H}(t)|| =  \int_{0}^{\tau}dt\sqrt{\mathrm{{Tr}}(\partial_t\hat{U}^\dagger \partial_t\hat{U})}\;.
\end{eqnarray}
The term $S_2$ is introduced to restrict the Hamiltonian $\hat{H}\in\mathcal{A}$, by requiring its orthogonality with $\hat{D} = \sum_{l} \lambda_l\hat{B}_l$
\begin{eqnarray}
    S_2 &=& \int_{0}^{\tau}dt\, \mathrm{Tr}\left(\hat{D}\hat{H}\right) = i\int_{0}^{\tau}dt\, \mathrm{Tr}\left(\hat{D}\partial_t\hat{U}\hat{U}^\dagger\right),
\end{eqnarray}
where $\hat{B}_l \in \mathcal{B}\equiv\mathcal{M/A}$ and $\lambda_l$ are Lagrange multipliers.
Varying $S_2$ with respect to Lagrange multipliers $\lambda_l$, we obtain $\mathrm{Tr}(\hat{B}_l\hat{H})=0$, which excludes any contributions beyond real nearest-neighbor couplings.

The contribution $S_3$ includes only boundary terms to restrict an optimal trajectory between initial and target states
\begin{eqnarray}
    S_3 &=& \mathrm{Tr}\left(\hat{R}_0(\hat{U} \hat{P}_0\hat{U}^\dagger-\hat{P}_0)\right)\delta(t)+\mathrm{Tr}\left(\hat{R}_1(\hat{U} \hat{P}_0\hat{U}^\dagger-\hat{P}_1)\right)\delta(t-\tau)\;,
\end{eqnarray}
where $\hat{P}_0 = \ket{\psi_0}\bra{\psi_0}$ and $\hat{P}_1 = \ket{\psi_1}\bra{\psi_1}$ are projections on initial $\ket{\psi_0}$ and target $\ket{\psi_1}$ states, and $\hat{R}_{0,1}$ are Lagrange multiplier matrices. The variation with respect to Lagrange multipliers $\hat{R}_{0}$ and $\hat{R}_{1}$ leads to $\hat{U}(0) \hat{P}_0\hat{U}^\dagger(0)-\hat{P}_0=0$ and $\hat{U}(\tau) \hat{P}_0\hat{U}^\dagger(\tau)-\hat{P}_1=0$, respectively. These conditions define the density matrix $\rho(t)=\ket{\psi(t)}\bra{\psi(t)} = \hat{U}\hat{P}_0\hat{U}^\dagger$ at the initial and final time $\rho(0) = \hat{P}_0$ and $\rho(\tau) = \hat{P}_1$, while the global phase of the wave function is discarded.

To derive quantum brachistochrone equations, we vary the cost functional with respect to the evolution operator $\hat{U}(t)$. The evolution operator  $\hat{U}(t)$ is unitary: $\hat{U}\,\hat{U}^\dagger=\hat{I}$. Hence, the variations $\delta\hat{U}$ and $\delta\hat{U}^\dagger$ are not independent, but rather connected to each other as
\begin{equation}
    \delta\hat{U}^\dagger=-\hat{U}^\dagger\,\delta\hat{U}\,\hat{U}^\dagger\:.
\end{equation}

With this, we compute the variation of the first term $S_1$ with respect to $\hat{U}$:
\begin{eqnarray}
    \delta S_1 
    %
    &=& \dfrac{i}{||\hat{H}||}\mathrm{{Tr}}(\hat{U}^\dagger \hat{H}\delta\hat{U})|_{0}^{\tau}
    -\dfrac{i}{||\hat{H}||}\int_{0}^{\tau}dt\mathrm{{Tr}}(
    \hat{U}^\dagger \partial_t \hat{H} \delta\hat{U}
    )
\end{eqnarray}
where we used that $||\hat{H}(t)||$ is constant along the trajectory. The variation of the second term $S_2$ yields
\begin{eqnarray}
    \delta S_2 
    &=& i \mathrm{Tr}
    (\hat{U}^\dagger \hat{D}\delta\hat{U})|_{0}^{\tau}
    -\int_{0}^{\tau}dt\mathrm{Tr}\left(\hat{U}^\dagger(i\partial_t\hat{D}
    -\left[\hat{H}, \hat{D}\right])\delta\hat{U}\right) 
\end{eqnarray}

Variation of the third term $S_3$ provides
\begin{eqnarray}
    \delta S_3 
    &=& \mathrm{Tr}\left(\hat{U}^\dagger(0)(\hat{P}_0\hat{R}_0 
    -\hat{R}_0\hat{P}_0)\delta\hat{U}(0)
    \right)
    +\mathrm{Tr}\left(
    \hat{U}^\dagger(\tau)(\hat{P}_1\hat{R}_1
    -\hat{R}_1\hat{P}_1)\delta\hat{U}(\tau)
    \right)
\end{eqnarray}

Thus, the variation of total action
\begin{eqnarray}
    \delta S &=& 
    -i\int_{0}^{\tau}dt\mathrm{{Tr}}\left(
    \hat{U}^\dagger\left( 
    \dfrac{\partial_t \hat{H}}{||\hat{H}||} + \partial_t\hat{D}
    +i\left[\hat{H}, \hat{D}\right]
    \right)\delta\hat{U}
    \right) \nonumber\\
    &&-i\mathrm{{Tr}}\left(
    \hat{U}^\dagger(0) \left(
    \dfrac{\hat{H}(0)}{||\hat{H}||}
    + \hat{D}(0)
    +i[\hat{P}_0,\hat{R}_0]
    \right)\delta\hat{U}(0)
    \right)\nonumber\\
    &&+i\mathrm{{Tr}}\left(
    \hat{U}^\dagger(\tau) \left(
    \dfrac{\hat{H}(\tau)}{||\hat{H}||}
    + \hat{D}(\tau)
    -i[\hat{P}_1,\hat{R}_1]
    \right)\delta\hat{U}(\tau)\right)
\end{eqnarray}

Now the extremum condition $\delta S = 0$ results in
\begin{eqnarray}
    \label{eqS:QB}
    \partial_t(\hat{H} + \hat{D})+i\left[\hat{H},\hat{D}\right] &=& 0\;,\\
    \label{eqS:QBb0}
    \hat{H}(0)+\hat{D}(0)+i\left[\hat{P}_0,\hat{R}_0\right] &=& 0 \;,\\
    \label{eqS:QBb1}
    \hat{H}(\tau)+ \hat{D}(\tau)-i\left[\hat{P}_1,\hat{R}_1\right] &=& 0 \;.
\end{eqnarray}
where we renormalized $\hat{D}\rightarrow \hat{D}/||\hat{H}||$ and $\hat{R}_{0,1}\rightarrow \hat{R}_{0,1}/||\hat{H}||$, since $||\hat{H}||$ is a constant along the optimal trajectory.

Note that the previous works~\cite{Carlini2006,Carlini2007,Wang2015,Malikis2024} did not introduce the term $S_3$ setting the variations $\delta U(0) = \delta U (\tau)$ to zero.  The equation~\eqref{eqS:QB} is known as {\it quantum brachistrochrone equation (QBE)}.

Further, we project equations~\eqref{eqS:QB}-\eqref{eqS:QBb1} onto the basis matrices in $(N^2-1)$-dimensional space of traceless Hermitian matrices. We thus obtain the evolution of the control parameters
\begin{eqnarray}
    \label{eqS:cA}
    \sqrt{2}\partial_t J_m+i\mathrm{Tr}\left(\hat{A}_m\left[\hat{H},\hat{D}\right]\right) &=& 0\;,\\
    \label{eqS:cB}
    \partial_t \lambda_l+i\mathrm{Tr}\left(\hat{B}_l\left[\hat{H},\hat{D}\right]\right) &=& 0\;,\\
    \label{eqS:cb0A}
    \sqrt{2}J_m(0)+i\mathrm{Tr}\left(\hat{A}_m\left[\hat{P}_0,\hat{R}_0\right]\right) &=& 0 \;,\\
    \label{eqS:cb0B}
    \lambda_l(0)+i\mathrm{Tr}\left(\hat{B}_l\left[\hat{P}_0,\hat{R}_0\right]\right) &=& 0 \;,\\
    \label{eqS:cb1A}
    \sqrt{2}J_m(\tau)-i\mathrm{Tr}\left(\hat{A}_m \left[\hat{P}_1,\hat{R}_1\right]\right) &=& 0 \;,\\
    \label{eqS:cb1B}
    \lambda_l(\tau)-i\mathrm{Tr}\left(\hat{B}_l \left[\hat{P}_1,\hat{R}_1\right]\right) &=& 0 \;,
\end{eqnarray}

\section{II. Equations for the single-particle transfer in a 1D array with time-varying nearest-neighbor couplings}

We consider $N$-dimensional single-particle sector of the Hilbert space corresponding to the array of $N$ qubits. In this sector, all operators are represented as $N\times N$ matrices. As a suitable basis, we choose generalized Gell-Mann matrices~\cite{Bertlmann2008}. In particular, we introduce the notation
\begin{equation}
  X_{mn}(z)=\frac{1}{\sqrt{2}}\,\left(z\,E_{nm}+z^*\,E_{mn}\right)\:,
\end{equation}
where $z$ is a complex number. $E_{nm}$ is a matrix with the elements $(E_{nm})_{pq}=\delta_{np}\delta_{mq}$, i.e. all its elements except one are equal to zero, and the remaining element with $(n,m)$ indices is equal to 1. It is straightforward to check that
\begin{equation}
    E_{ij}\,E_{kl}=E_{il}\,\delta_{jk}\:.
\end{equation}


The subspace $\mathcal{A}$ describes nearest-neighbor couplings between the qubits and contains $N-1$ symmetric element of $\mathcal{M}$:
\begin{eqnarray}\label{eqs:Amatrix}
    A_j &=& X_{j,j+1}(1) \equiv (E_{j+1,j}+E_{j,j+1})/\sqrt{2}\;,\; 1\leq j \leq N-1\;.
\end{eqnarray}
Next we introduce the matrices
\begin{equation}\label{eqs:Bmatrix}
 B_{m,m+q}^{(e)}=X_{m,m+q}(i^{q-1})
\end{equation}
with $2\leq q\leq N-m$. Overall, there are $(N-1)(N-2)/2$ of such matrices. Besides that, there are $(N-1)(N+2)/2$ of other matrices belonging to the $\mathcal{B}$ subspace. However, as we demonstrate below, they do not appear in quantum brachistochrone equations being orthogonal to the introduced set. Therefore we do not specify their explicit form.


Having chosen a specific basis, we analyze the equations~\eqref{eqS:cA}-\eqref{eqS:cb1B}. We denote by $\lambda_{m,m+q}$ Lagrange coefficients which appear in front of the matrices $B_{m,m+q}^{(e)}$. $\mathcal{X}$ denotes the subspace spanned by $A_m$ and $B_{m,m+q}^{(e)}$ matrices.

Since we aim to transfer a single excitation from the leftmost qubit $\ket{\psi_0}=\ket{1_1}$ to the rightmost one $\ket{\psi_1}=\ket{1_N}$, the projectors $\hat{P}_0 = E_{1,1}$ and $\hat{P}_1 = E_{N,N}$.

Moreover, the commutators of $X_{n,n+p}(i^{p-1})$ matrices can be readily computed using the identity
\begin{eqnarray}
    i\sqrt{2}\left[X_{n,n+p}(i^{p-1}),X_{m,m+q}(i^{q-1})\right] 
    &=& X_{n-q,n+p}(i^{p+q-1})\delta_{n,m+q} -X_{n,n+p+q}(i^{p+q-1})\delta_{n+p,m}  \nonumber\\
    && +X_{n+q,n+p}(i^{p-q+1}) \delta_{n,m} 
    -
    X_{n,n+p-q}(i^{p-q+1})\delta_{n+p,m+q}
\end{eqnarray}

Starting from Eq.~\eqref{eqS:cA}, we note that $\hat{A}_m = \hat{X}_{m,m+1}(1)\in\mathcal{A}\in\mathcal{X}$, leading to

\begin{eqnarray}
    \partial_t J_m&=&-i\sum_{l,k}\lambda_l J_k\mathrm{Tr}\left(\hat{B}_l\left[\hat{A}_m,\hat{A}_k\right]\right)\;,\\
    \partial_t J_m &=&-\dfrac{1}{\sqrt{2}}\sum_{l,k}\lambda_l J_k\mathrm{Tr}\left(\hat{B}_l
    X_{m-1,m+1}(i)\delta_{m,k+1} -\hat{B}_lX_{m,m+2}(i)\delta_{m+1,k}
    \right)\;,\\
    \partial_t J_m &=&\dfrac{1}{\sqrt{2}}\left(J_{m+1}\lambda_{m,m+2} - J_{m-1}\lambda_{m-1,m+1}
    \right)\;.
\end{eqnarray}

Since coupling depends only on $\lambda_{m,m+2}$, in equation~\eqref{eqS:cB} we can consider only elements from  $\mathcal{X}$

\begin{eqnarray}
    \partial_t \lambda_{n,n+p}&=&-i\sqrt{2}\sum_{m,l}\lambda_lJ_m\mathrm{Tr}\left(\hat{B}_l\left[\hat{X}_{n,n+p}(i^{p-1}),\hat{X}_{m,m+1}(1)\right]\right) \;,\\
    \partial_t \lambda_{n,n+p}&=&
    -\sum_{m,l}\lambda_lJ_m\mathrm{Tr}\left(
    \hat{B}_l X_{n-1,n+p}(i^{p})\delta_{n,m+1} 
    -\hat{B}_l X_{n,n+p+1}(i^{p})\delta_{n+p,m}  
    \right)
    \nonumber\\&&
    -\sum_{m,l}\lambda_lJ_m\mathrm{Tr}\left(
    \hat{B}_l X_{n+1,n+p}(i^{p}) \delta_{n,m} 
    - \hat{B}_l X_{n,n+p-1}(i^{p})\delta_{n+p,m+1}
    \right)\nonumber\\
    &=&
    -\sum_{m,l}\lambda_lJ_m\mathrm{Tr}\left(
    \hat{B}_l X_{n-1,n-1+p+1}(i^{p+1-1})\delta_{n,m+1} 
    -\hat{B}_l X_{n,n+p+1}(i^{p+1-1})\delta_{n+p,m}  
    \right)
    \nonumber\\&&
    +\sum_{m,l}\lambda_lJ_m\mathrm{Tr}\left(
    \hat{B}_l X_{n+1,n+1+p-1}(i^{p-2}) \delta_{n,m} 
    - \hat{B}_l X_{n,n+p-1}(i^{p-1})\delta_{n+p,m+1}
    \right)
\end{eqnarray}

\begin{eqnarray}
    \partial_t \lambda_{n,n+p}&=&
    J_{n+p}\lambda_{n,n+p+1}
    -J_{n-1}\lambda_{n-1,n+p}
    - J_{n+p-1}\lambda_{n,n+p-1}
    +J_{n}\lambda_{n+1,n+p}
\end{eqnarray}

Initial conditions~\eqref{eqS:cb0A}-\eqref{eqS:cb0B} at $t=0$ read

\begin{eqnarray}
    \sqrt{2}J_m(0)&=&-\dfrac{i}{\sqrt{2}}\mathrm{Tr}\left((E_{m,m+1}E_{1,1}+E_{m+1,m}E_{1,1}-E_{1,1}E_{m,m+1}-E_{1,1}E_{m+1,m})\hat{R}_0\right) \;,\\
    \lambda_{n,n+p}(0)&=&-\dfrac{i}{\sqrt{2}}\mathrm{Tr}\left((i^{p-1}E_{n+p,n}+(-i)^{p-1}E_{n,n+p})E_{1,1}-E_{1,1}(i^{p-1}E_{n+p,n}+(-i)^{p-1}E_{n,n+p}))\hat{R}_0\right) \;,
\end{eqnarray}

\begin{eqnarray}
    \sqrt{2}J_m(0)&=&-\dfrac{i}{\sqrt{2}}\mathrm{Tr}\left(((E_{0,1}\delta_{m,0}+E_{2,1}\delta_{m,1})-(E_{1,2}\delta_{m,1}+E_{1,0}\delta_{m,0}))\hat{R}_0\right) \;,\\
    \lambda_{n,n+p}(0)&=&-\dfrac{i}{\sqrt{2}}\mathrm{Tr}\left((i^{p-1}E_{1+p,1}\delta_{n,1}+(-i)^{p-1}E_{1-p,1}\delta_{n+p,1}-i^{p-1}E_{1,1-p}\delta_{1,n+p}-(-i)^{p-1}E_{1,1+p}\delta_{1,n}))\hat{R}_0\right)  \;,
\end{eqnarray}

\begin{eqnarray}
    \sqrt{2}J_m(0)&=&-\delta_{m,1}\dfrac{i}{\sqrt{2}}\sum_{k,k'}\left((E_{2,1})_{kk'}(\hat{R}_0)_{k'k}-(E_{1,2})_{kk'}(\hat{R}_0)_{k'k}\right) = i\delta_{m,1}((\hat{R}_0)_{21}-(\hat{R}_0)_{12})/\sqrt{2} \;,\\
    \lambda_{n,n+p}(0)&=&-\delta_{n,1}\dfrac{i}{\sqrt{2}}\sum_{k,k'}\left(i^{p-1}(E_{1+p,1})_{kk'}(\hat{R}_0)_{k'k}-(-i)^{p-1}(E_{1,1+p})_{kk'}(\hat{R}_0)_{k'k}\right)\nonumber\\
    &=&i\delta_{n,1}\left((-i)^{p-1}(\hat{R}_0)_{1+p,1}-i^{p-1}(\hat{R}_0)_{1,1+p}\right)/\sqrt{2}
\end{eqnarray}

We thus conclude that $J_2(0)=\dots=J_{N-1}(0)=0$ and the only nonzero coupling at $t=0$ is $J_1$. In the same way, we recover that the only nonzero Lagrange multipliers at $t=0$ are $\lambda_{1,1+p}$ for $2\leq p\leq N-1$. This provides much-needed initial conditions to solve quantum brachistochrone equations, overall $(N-1)(N-2)/2$ initial conditions. 

Boundary conditions~\eqref{eqS:cb0A}-\eqref{eqS:cb0B} at the moment of time $t=\tau$ yield

\begin{eqnarray}
    \sqrt{2}J_m(\tau)&=&-\dfrac{i}{\sqrt{2}}\mathrm{Tr}\left((E_{m,m+1}E_{N,N}+E_{m+1,m}E_{N,N}-E_{N,N}E_{m,m+1}-E_{N,N}E_{m+1,m})\hat{R}_0\right) \;,\\
    \lambda_{n,n+p}(\tau)&=&-\dfrac{i}{\sqrt{2}}\mathrm{Tr}\left((i^{p-1}E_{n+p,n}+(-i)^{p-1}E_{n,n+p})E_{N,N}-E_{N,N}(i^{p-1}E_{n+p,n}+(-i)^{p-1}E_{n,n+p}))\hat{R}_0\right) \;,
\end{eqnarray}

\begin{eqnarray}
    \sqrt{2}J_m(\tau)&=&-\dfrac{i}{\sqrt{2}}\mathrm{Tr}\left((E_{N-1,N}\delta_{m,N-1}-E_{N,N-1}\delta_{m,N-1})\hat{R}_0\right) \;,\\
    \lambda_{n,n+p}(\tau)&=&-\dfrac{i}{\sqrt{2}}\mathrm{Tr}\left(((-i)^{p-1}E_{N-p,N}\delta_{n+p,N}-i^{p-1}E_{N,N-p}\delta_{n+p,N})\hat{R}_0\right)  \;,
\end{eqnarray}

\begin{eqnarray}
    \sqrt{2}J_m(\tau)&=&-\delta_{m,N-1}\dfrac{i}{\sqrt{2}}\sum_{k,k'}\left((E_{N-1,N})_{kk'}(\hat{R}_0)_{k'k}-(E_{N,N-1})_{kk'}(\hat{R}_0)_{k'k}\right) \nonumber\\&=& i\delta_{m,N-1}((\hat{R}_0)_{N,N-1}-(\hat{R}_0)_{N-1,N})/\sqrt{2} \;,\\
    \lambda_{n,n+p}(\tau)&=&\delta_{n+p,N}\dfrac{i}{\sqrt{2}}\sum_{k,k'}\left(i^{p-1}(E_{N,N-p})_{kk'}(\hat{R}_0)_{k'k}-(-i)^{p-1}(E_{N-p,N})_{kk'}(\hat{R}_0)_{k'k}\right)\nonumber\\
    &=&i\delta_{n+p,N}\left(i^{p-1}(\hat{R}_0)_{N-p,N}-(-i)^{p-1}(\hat{R}_0)_{N,N-p}\right)/\sqrt{2}    
\end{eqnarray}

Hence, the only nonzero coupling at $t=\tau$ is $J_{N-1}(\tau)$, while $J_1(\tau)=\dots=J_{N-2}(\tau)=0$. The only nonzero Lagrange multipliers at the final moment of time are $\lambda_{N-p,N}$, where $2\leq p\leq (N-1)$. This provides the boundary conditions for quantum brachistochrone problem, overall  $(N-1)(N-2)/2$ boundary conditions. 







\begin{table}[h]
    \centering
    \begin{tabular}{|c|c|c|c|c|c|c|c|c|}
    \hline
        N & 3 & 4 & 5 & 6 & 7 & 8 & 9 & 10\\\hline
        $\tau$ & 2.7207 & 3.85444 & 4.98542 & 6.11586  & 7.2462 & 8.37651 & 9.50682  & 10.6371 \\\hline
        $\lambda_{1,3}$ & -0.816497 & -0.869945 & -0.879405 & -0.881276 & -0.881655 & -0.881733 & -0.881749 & -0.881752 \\\hline
        $\lambda_{1,4}$ & - & 0.743041 & 0.800224 & 0.810655 & 0.812732 & 0.813154 & 0.81324 & 0.813258 \\\hline
        $\lambda_{1,5}$ & - & - & -0.727694 & -0.785265 & -0.795828 & -0.797934 & -0.798363 &  -0.79845 \\\hline
        $\lambda_{1,6}$ & - & - & - & 0.724395 & 0.782011 & 0.792594 & 0.794705 & 0.795135 \\\hline
        $\lambda_{1,7}$ & - & - & - & - & -0.723702 &  -0.781325 & -0.791911 & -0.794023 \\\hline
        $\lambda_{1,8}$ & - & - & - & - & - & 0.723559 & 0.781183 & 0.79177 \\\hline
        $\lambda_{1,9}$ & - & - & - & - & - & - & -0.723529  & -0.781153 \\\hline 
        $\lambda_{1,10}$ & - & - & - & - & - & - & - & 0.723523 \\\hline 
    \end{tabular}
    \caption{Calculated transfer time $\tau$ and initial values of the coefficients $\lambda_{1,2+n}(0)$, $1\leq n\leq N-2$ for the arrays of qubits with the length $N\leq 10$ with $J_0=1$. }
    \label{tab:my_label}
\end{table}

\begin{table}[]
    \centering
    \begin{tabular}{|c|c|c|c|c|c|c|c|c|}
    \hline
        N & 20 & 25 & 30 & 35 & 70 & 75 & 80 & 100\\\hline
        $\tau$ & 21.9402 & 27.5917 & 33.2433 & 38.8948 & 78.4555 & 84.1163 & 89.7586 & 112.3773  \\\hline
    \end{tabular}
    \caption{The transfer time $\tau$ obtained for longer qubit arrays $N\leq100$ with numerical gradient-based optimization of the target state fidelity ($F_{N\leq80} =0.9999999998$). The optimal time for $N=100$ is evaluated from Eq.~\eqref{eq:Scaling} and provides fidelity $F_{100} = 0.9998$.  $J_0=1$.}
    \label{tab:T2}
\end{table}

\begin{figure}[h]
   \begin{minipage}{0.48\textwidth}
     \centering
     \includegraphics[width=1\linewidth]{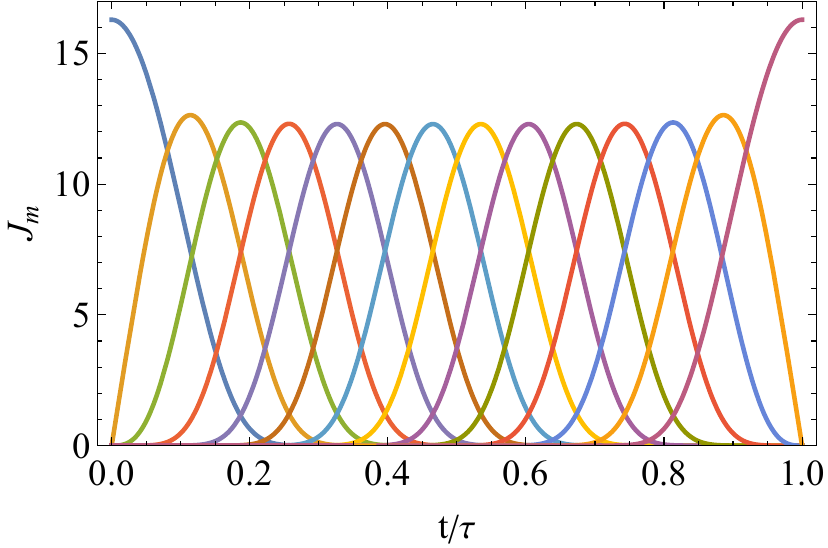}
     \caption{Temporal dependence of the couplings $J_m$ for all sites in a 15-qubit array enabling time-optimal transfer. $J_0=1$.}\label{psi15}
   \end{minipage}\hfill
   \begin{minipage}{0.48\textwidth}
     \centering
     \includegraphics[width=1\linewidth]{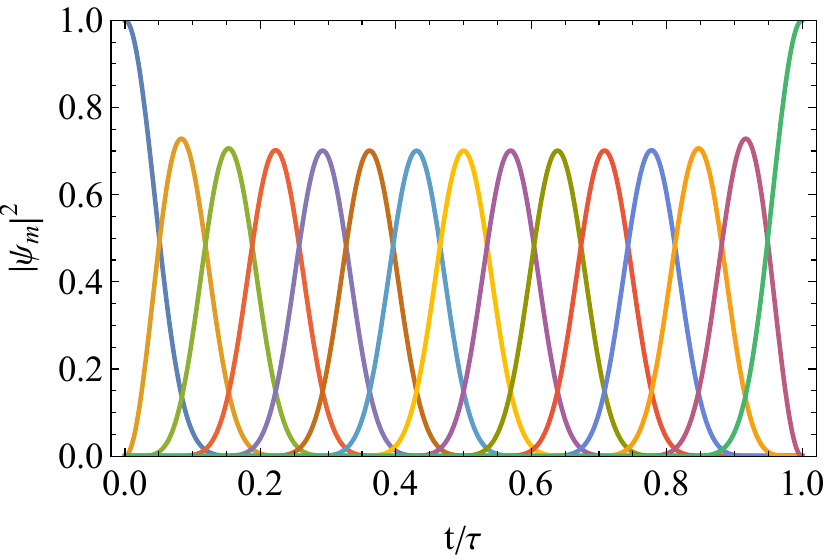}
     \caption{Associated probability distributions $|\psi_m(t)|^2$ at all sites of a 15-qubit array.}\label{J15}
   \end{minipage}
\end{figure}

\end{document}